\documentclass[conference]{IEEEtran}
\usepackage{color}
\usepackage[english]{babel}
\usepackage{graphicx}
\usepackage{epstopdf}
\usepackage{times}
\usepackage{multirow}
\usepackage[algo2e,linesnumbered]{algorithm2e} 
\usepackage{algorithmic}
\usepackage{algorithm}
\usepackage{mathenv}
\usepackage{array}
\usepackage[T1]{fontenc}
\usepackage{array}
\usepackage{mdwmath}
\usepackage{multicol}
\usepackage{amsmath}
\usepackage{amssymb}
\usepackage{float}
\usepackage{url}
\newcommand{\subparagraph}{}
\usepackage{lipsum}
\usepackage{titlesec}
\titlespacing\section{0pt}{6pt plus 2pt minus 2pt}{6pt plus 2pt minus 2pt}
\titlespacing\subsection{0pt}{3pt plus 2pt minus 2pt}{3pt plus 2pt minus 2pt}

\usepackage{mathtools}

\usepackage{dirtytalk}
\usepackage[utf8]{inputenc}
\usepackage{amsthm}
\newtheorem{theorem}{Theorem}
\newtheorem{definition}{Definition}
\makeatletter
\newcommand*{\rom}[1]{\expandafter\@slowromancap\romannumeral #1@}
\makeatother

\showoutput
\begin{document}

\long\def\/*#1*/{}
\setlength{\abovedisplayskip}{2pt}
\setlength{\belowdisplayskip}{2pt}

\title{Distributed Scheme for Interference Mitigation of \textit{WBAN}s Using Predictable Channel Hopping}

  \author{\IEEEauthorblockN{Mohamad Ali\IEEEauthorrefmark{1}, Hassine Moungla\IEEEauthorrefmark{1}, Mohamed Younis\IEEEauthorrefmark{2}, Ahmed Mehaoua\IEEEauthorrefmark{1}}
\IEEEauthorblockA{\IEEEauthorrefmark{1}LIPADE, University of Paris Descartes, Sorbonne Paris Cit\'{e}, Paris, France \\
}{\IEEEauthorrefmark{2}Department of Computer Science and Electrical Engineering, University of Maryland, Baltimore County, United States} \\Email: \{mohamad.ali; hassine.moungla; ahmed.mehaoua\}@parisdescartes.fr; younis@umbc.edu}
\maketitle
\begin{abstract}
When sensors of different coexisting wireless body area networks (\textit{WBAN}s) transmit at the same time using the same channel, a co-channel interference is experienced and hence the performance of the involved \textit{WBAN}s may be degraded. In this paper, we exploit the \textit{16 channels} available in the \textit{2.4 GHz} international band of \textit{ZIGBEE}, and propose a distributed scheme that avoids interference through predictable channel hopping based on Latin rectangles, namely, \textit{CHIM}. In the proposed \textit{CHIM} scheme, each \textit{WBAN}'s coordinator picks a Latin rectangle whose rows are \textit{ZIGBEE} channels and columns are sensor \textit{IDs}. Based on the Latin rectangle of the individual \textit{WBAN}, each sensor is allocated a backup time-slot and a channel to use if it experiences interference such that collisions among different transmissions of coexisting \textit{WBAN}s are minimized. We further present a mathematical analysis that derives the collision probability of each sensor's transmission in the network. In addition, the efficiency of \textit{CHIM} in terms of transmission delay and energy consumption minimization are validated by simulations.
\end{abstract}
%\keywords{{\small{\textit{WBAN}s Coexistence,}\small{\textit{ZIGBEE} Channels,}\small{Interference Mitigation,}\small{Latin Rectangles} }}

\section{Introduction}
A \textit{WBAN} is a wireless short range communication network comprises of a single coordinator and multiple low power and wearable sensors for collecting the human personal data. Example applications of \textit{WBAN}s such as ubiquitous health care, medical treatment, consumer electronics, sports and military \cite{key15}. For example, the involved sensors may be observing the heart and the brain electrical activities as well as blood pressure, core temperature, oxygen saturation, carbon dioxide concentration, etc. %\textcolor{red}{\textit{WBAN}s aim to improve the speed, accuracy and the reliability of sensors communication that are distributed on, within and in the immediate proximity of a human body.}

Recently, \textit{ZIGBEE} standard \cite{key26} has proposed new specifications for the physical and medium access layers and determined an upper bound for the number of \textit{WBAN}s and sensors that may collocate within a certain communication range. Thus, there is great possibility of inter-\textit{WBAN} interference in environments such as hospitals and senior communities, where \textit{WBAN}s are densely deployed. Consequently, the interference may require multiple retransmissions or even hinder the reception of the data and thus degrade the performance of each individual \textit{WBAN} as whole. Therefore, interference mitigation is quite necessary to avoid repeated transmissions and data loss and hence increase maximum network lifetime as well as reduce the minimum delay and unnecessary communication related energy consumption. To this end, three mechanisms, namely, beacon shifting, channel hopping and active superframe interleaving are proposed for \textit{WBAN}s interference mitigation in \textit{ZIGBEE} standard \cite{key26}.

In addition, the co-channel interference is challenging due to the highly mobile and resource constrained nature of \textit{WBAN}s\/*\textcolor{red}{(e.g. limited energy resource, communication capabilities, memory and computation, etc)}*/. The uncontrolled motion pattern and the independent operation of \textit{WBAN} make the interference mitigation by a centralized unit as well as the application of advanced communication and power control techniques used in other wireless networks, unsuitable for \textit{WBAN}s. Though, \textit{ZIGBEE} standard \cite{key26} has recommended the use of \textit{TDMA} medium access scheme as an alternative solution to avoid intra-\textit{WBAN} co-channel interference, nonetheless and due to the absence of coordination and synchronization among \textit{WBAN}s, the different superframes may overlap and the concurrent transmissions of different nearby \textit{WBAN}s may still interfere. More specifically, when two or more sensors of different \textit{WBAN}s access the shared channel at the same time, their transmissions cause medium access collision. This paper tackles these issues and contributes the following:
\begin{itemize}
 \item  \textit{CHIM, a distributed scheme that enables predictable channel hopping using Latin rectangles in order to avoid interference among coexisting \textit{WBAN}s}
 \item \textit{An analysis of the collision probability model for sensors transmissions}
\end{itemize}
The simulation results and theoretical analysis show that our approach can significantly lower the number of collisions and reduce the delay among the individual transmissions of coexisting \textit{WBAN}s as well as increase the energy savings at \textit{sensor}- and \textit{WBAN}-levels.\/*\textcolor{red}{Moreover, \textit{CHIM} significantly avoids the inter-\textit{WBAN} interference and do not require any mutual coordination among the individual coordinators.}*/ The rest of the paper is organized as follows. Section \rom{2} sets our work apart from other approaches in the literature. Section \rom{3} summarizes the system model and provides a brief overview of Latin squares. Section \rom{4} describes \textit{CHIM} in detail. Section \rom{5} analyzes the collision probability of \textit{CHIM}. Section \rom{6} presents the simulation results. Finally the paper is concluded in Section \rom{7}.

\section{Related Work}
Avoidance and mitigation of channel interference has been extensively researched in the wireless communication literature. Published techniques in the realm of \textit{WBAN} can be categorized as spectrum allocation, cooperative communication, power control and multiple medium access schemes. Example schemes that pursue the spectrum allocation methodology include \cite{key16,key25,key1}. Movassaghi et al., \cite{key16} have proposed a distributed channel allocation for the sensors belonging to interference regions amongst coexisting \textit{WBAN}s. Whereas, in \cite{key25}, an adaptive scheme that allocates synchronous and parallel transmission intervals has been proposed for \textit{sensor}-level interference avoidance rather than considering each \textit{WBAN} as whole. Moreover, this scheme is optimized \/*\textcolor{red}{considers different quality of service parameters}*/to reduce the number of orthogonal channels\/*\textcolor{red}{and achieve better usage of limited resources in \textit{WBAN}s}*/. Meanwhile, Movassaghi et al., \cite{key1} have also proposed an algorithm for dynamic channel allocation amongst coexisting \textit{WBAN}s, where variations in channel assignment due to \textit{WBAN} mobility scenarios are investigated. \/*Further, the scheme is optimized to allocate transmission time-slots with synchronous and parallel transmissions such that the interference is avoided and hence and to reduce the number of interfering sensors is reduced.*/

\/*\textcolor{red}{A number of approaches have pursued cooperative communication and power control to mitigate co-channel interference.}*/ Meanwhile, Dong et al., \cite{key9} have adopted cooperative communication integrated with transmit power control for multiple coexisting \textit{WBAN}s. \/*Similarly, Ali et al., \cite{key54}, have proposed a distributed, carrier sense multiple access with collision avoidance scheme that enables interference-free sensors to transmit through a priori-agreed upon channel. Whilst, interfering sensors may double their contention windows or use a switched channel to avoid intra-\textit{WBAN} interference*/ Whereas, Zou et al., \cite{key30} have proposed a \textit{Bayesian} game based power control approach to mitigate the impact of inter-\textit{WBAN} interference. 

Other approaches have pursued multiple medium access schemes for interference mitigation. Kim et al., \cite{key21} have pursued multiple medium access schemes and proposed a distributed \textit{TDMA}-based beacon interval shifting scheme for avoiding the overlap between superframe's active period through employing carrier sense before a beacon transmission. Similarly, Chen et al., \cite{key3} have adopted \textit{TDMA} for scheduling intra-\textit{WBAN} transmissions and carrier sensing to deal with inter-\textit{WBAN} interference. Meanwhile the approach of \cite{key57} has mapped the channel allocation as a graph coloring problem. The coordinators need to exchange messages to achieve a non-conflict coloring in a distributed manner. Whilst, Ju et al., \cite{key53} have proposed a multi-channel topology-transparent algorithm based on Latin squares for transmissions scheduling in multihop packet radio networks. Thus, in a multi-channel \textit{TDMA}-based network, each node is equipped with a single transmitter and multiple receivers. Like \cite{key53}, \textit{CHIM} employs Latin rectangles to form a predictable non-interfering transmission schedule. However, \textit{CHIM} considers the presence of single receiver rather than multiple receivers per a single node and single hop rather than multihop communication. In addition, \textit{CHIM} avoids frequent channel switching by limiting it to the case when a sensor interference occurs. 

Unlike prior work, in this paper we exploit the \textit{16} channels available in the \textit{2.4 GHz}  international band of \textit{ZIGBEE}, and propose a distributed scheme based on predictable channel hopping for interference avoidance amongst coexisting \textit{WBAN}s. At the network setup, each individual \textit{WBAN} autonomously picks a Latin rectangle\/*\textcolor{red}{ whose rows are the \textit{ZIGBEE} channels and columns are the sensor IDs. Based on the Latin rectangle of the individual \textit{WBAN},}*/ through which each sensor is allocated a backup time-slot and channel to use if it experiences interference. We depend on the special properties of Latin rectangles to minimize the probability of both time and channel matching among sensors in different \textit{WBAN}s, and consequently reduce the transmission delay and energy consumption.

\section{System Model and Preliminaries}
\subsection{System Model and Assumptions}
We consider \textit{N} \textit{TDMA}-based \textit{WBAN}s that coexist  in an operation area, e.g., \/*\textcolor{red}{when a group of patients moving around in}*/a large hall of a hospital. Each \textit{WBAN} consists of a single coordinator denoted by \textit{Crd} and up to \textit{K} sensors, each transmits its data at maximum rate of \textit{250Kb/s} within the \textit{2.4 GHz} international band (\textit{ISM}). Furthermore, we assume all coordinators are equipped with significantly richer energy supply than sensors and all sensors have access to all \textit{ZIGBEE} channels at any time.

\/*\textcolor{red}{Due to the \textit{WBAN}'s unpredictable motion pattern, it is very hard to achieve inter-\textit{WBAN}s coordination or to have a central unit to mitigate the potential interference when some of them are in close proximity of each other.}*/ Basically, co-channel interference may arise due to the collisions amongst the concurrent transmissions made by sensors in different \textit{WBAN}s in the same time-slot (\textit{TS}). To address this issue, we exploit the \textit{16 channels} in the \textit{2.4 GHz} \textit{ISM} band of \textit{ZIGBEE} to resolve this problem through predicatable channel hopping. However, to avoid the delay and energy overhead due to frequent change in channels, our approach enables hopping only at the level of the interfering sensors.
\subsection{Latin Squares}
In this section, we provide a brief overview of Latin squares that we used to allocate interference mitigation channels.
\begin{definition}
A Latin square is a $K \times K$ matrix, filled with \textit{K} distinct symbols, each symbol appearing once in each column and once in each row.
\end{definition}
\begin{definition}\label{orthogonal}
Two distinct $K \times K$ Latin squares \textit{E = ($e_{i,j}$)} and \textit{F = ($f_{i,j}$)}, so that $e_{i,j}$ and $f_{i,j}$ $\in$ $\{1,2, \dots K\}$, are said to be orthogonal, if the $K^{2}$ ordered pairs ($e_{i,j},f_{i,j}$) are all different from each other. More generally, the set \textit{$O=\{E_{1}, E_{2}, E_{3},\dots,E_{r}\}$} of distinct Latin squares \textit{E} is said to be orthogonal, if every pair in \textit{O} is orthogonal.
\end{definition}
\begin{definition}
An orthogonal set of Latin squares of order \textit{K} is of size \textit{(K-1)}, i.e., the number of Latin squares in the orthogonal family is \textit{(K-1)}, is called a complete set \cite{key52, key53}.
\end{definition}
\begin{definition}
A $M \times K$ Latin rectangle is a $M \times K$ matrix \textit{G}, filled with symbols $a_{ij}$ $\in$ $\{1,2,\dots,K\}$, such that each row and each column contains only distinct symbols.
\end{definition}
\begin{theorem}\label{theo1}
If there is an orthogonal family of \textit{r} Latin squares of order \textit{K}, then \textit{$r\leq K-1$} \cite{key52}
\end{theorem}
\textit{E} and \textit{F} are orthogonal Latin squares of order \textit{3}, because no two ordered pairs within \textit{E$ \bowtie$ F} are similar.
\begingroup\makeatletter\def\f@size{8}\check@mathfonts
\begin{center}
$E = 
\begin{bmatrix}
   1&2&3 \\
   2&3&1 \\
   3&1&2 \\
\end{bmatrix}$ $F = 
\begin{bmatrix}
   1&2&3 \\
   3&1&2 \\
   2&3&1 \\
\end{bmatrix}$
$E\bowtie F = 
\begin{bmatrix}
   1,1&2,2&3,3\\
   2,3&3,1&1,2\\
   3,2&1,3&2,1\\
\end{bmatrix}$
\end{center}
\endgroup
Basically, if a \textit{WBAN} picks one Latin square from an orthogonal set, there will be no shared channel among the coexisting Latins. According to \textit{\textbf{Theorem \ref{theo1}}}, the number of \textit{WBAN}s using orthogonal Latin squares is upper bounded by \textit{K-1}, thus, \textit{K} should be large enough so that, each \textit{WBAN} can pick an orthogonal Latin square with high probability. The Latin square size will depend on the largest among the number of channels, denoted by \textit{M}, and number of sensors in each \textit{WBAN}, denoted by \textit{K}. The \textit{ZIGBEE} standard \cite{key26} limits the number of channels which constitutes the rows in the Latin square to \textit{16}, no more than \textit{16} transmissions can be scheduled. To overcome such a limitation, \textit{CHIM} employs Latin rectangles instead, i.e., does not restrict the value of \textit{K} and hence supports \textit{$K > M$}.

Throughout this paper, we denote a symbol by the ordered pair \textit{(i,j)} referenced at the $i^{th}$ row and $j^{th}$ column in the Latin rectangle, which refers to the assignment of $i^{th}$ interference mitigation backup channel, denoted by \textit{BKC}, to the $j^{th}$ sensor in the dedicated backup time-slot, denoted by \textit{BKTS}. In addition, \textit{C} and \textit{DFC}, respectively, denote a channel and default operation channel.
\section{Channel Hopping for Interference Mitigation}
As pointed out, a co-channel interference takes place if the simultaneous transmissions of sensors and coordinators in different \textit{WBAN}s collide. The potential for such a collision problem grows with the increase in the communication range and the density of sensors in the individual \textit{WBAN}s. To mitigate interference, \textit{CHIM} exploits the availability of multiple channels to assign each \textit{WBAN} a distinct default channel and in case of interference it allows the individual sensors to hop among the remaining channels in a pattern that is predictable within a \textit{WBAN} and random to the other coexisting \textit{WBAN}s. To achieve that, \textit{CHIM} extends the size of the superframe through the addition of extra interference mitigation backup time-slots and employs Latin rectangles as the underlying scheme for channel allocation to sensors. \textit{CHIM} relies on the properties of Latin rectangles in order to reduce the probability of collision while enabling autonomous scheduling of the medium access.
\subsection{Superframe Structure}
\begin{figure}
  \centering
       \includegraphics[width=0.3\textheight]{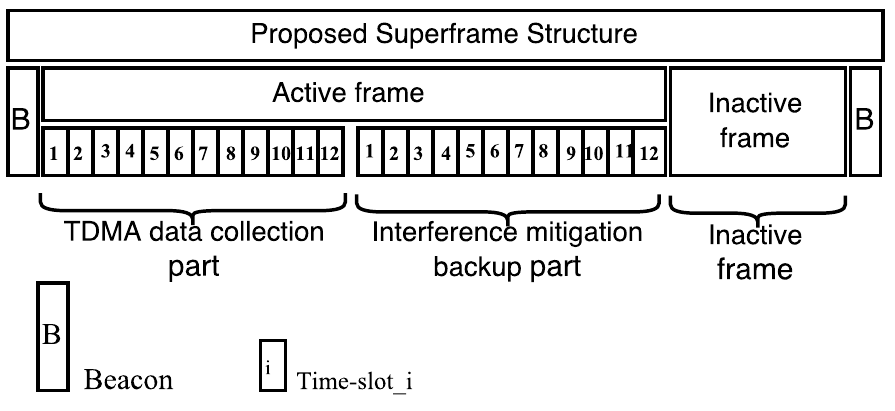}
\caption{Proposed superframe structure}
\label{superframestruct}
\end{figure}

We consider beacon-enabled \textit{WBAN}s, where each superframe is delimited by two beacons and composed of two successive frames: (i) active, that is dedicated for sensors, and (ii) inactive, that is designated for coordinators. The active frame is further divided into two parts of equal size, the time division multiple access (\textit{TDMA}) data-collection part and the interference mitigation backup (\textit{IMB}) interference mitigation part, each is of \textit{K} time-slots length. \textbf{Figure \ref{superframestruct}} shows the superframe structure.
In the \textit{TDMA} part, each sensor transmits its data packet in its assigned time-slot to the coordinator through the default channel. However, in the \textit{IMB} interference mitigation part, each interfering sensor retransmits the same data packet in its allocated backup time-slot to the coordinator through a priori-agreed upon channel. In interference-free conditions, the coordinator stays tuned to the default channel. If communication with a specific sensor $S_{i}$ fails during $S_{i}$'s designated time-slot, the coordinator will tune to the $S_{i}$'s backup channel during $S_{i}$'s time-slot in the \textit{IMB} interference mitigation part of the active frame. Whereas, during the inactive frame, all the sensors sleep and hence, the coordinators may transmit all data to a command center. 

We still need to determine the length of each frame. \/*\textcolor{red}{Each sensor within each \textit{WBAN} may require \textit{p} time-slots to complete its data transmission. For example, for a sensor that samples at a rate of \textit{10 per second}, we need \textit{10} time-slots in a frame of \textit{1 second}. If all sensors have the same requirement, \textit{$p \times K$} time-slots for a \textit{WBAN} of \textit{K} sensors are required in each \textit{TDMA} data collection part of the active frame.}*/ In fact, the size of the \textit{TDMA} data collection part depends on two factors, 1) how big the time-slot, which is based on the protocol in use, and 2) the number of required time-slots, which is determined by the different sampling rates of \textit{WBAN} sensors. Generally, the sum of number of samples for all sensors in a time period determines the \textit{TDMA} data collection part size. However, \textit{CHIM} requires the \textit{TDMA} data collection part for all WBANs to be the same length so that collision could be better avoided by unifying the frame size across the various \textit{WBAN} and leveraging the properties of Latin rectangles.\/*\textcolor{red}{ as we explain later late in this section. Therefore, in \textit{CHIM}, the \textit{TDMA} data collection part length is determined based on the highest sampling rate. For example, when \textit{p=1}, the number of time-slots to be made in the active frame is \textit{2 $\times$ K} time-slots, i.e., \textit{K} time-slots are for the \textit{TDMA} data-collection part and \textit{K} time-slots are for the \textit{IMB} interference mitigation backup part.}*/ Whereas, the inactive frame directly follows the active frame (\textit{TDMA} part and the \textit{IMB} part) and whose length depends on the underlying duty cycle scheme of the sensors.

\subsection{Collision Scenarios}
\begin{figure}
  \centering
       \includegraphics[width=0.25\textheight]{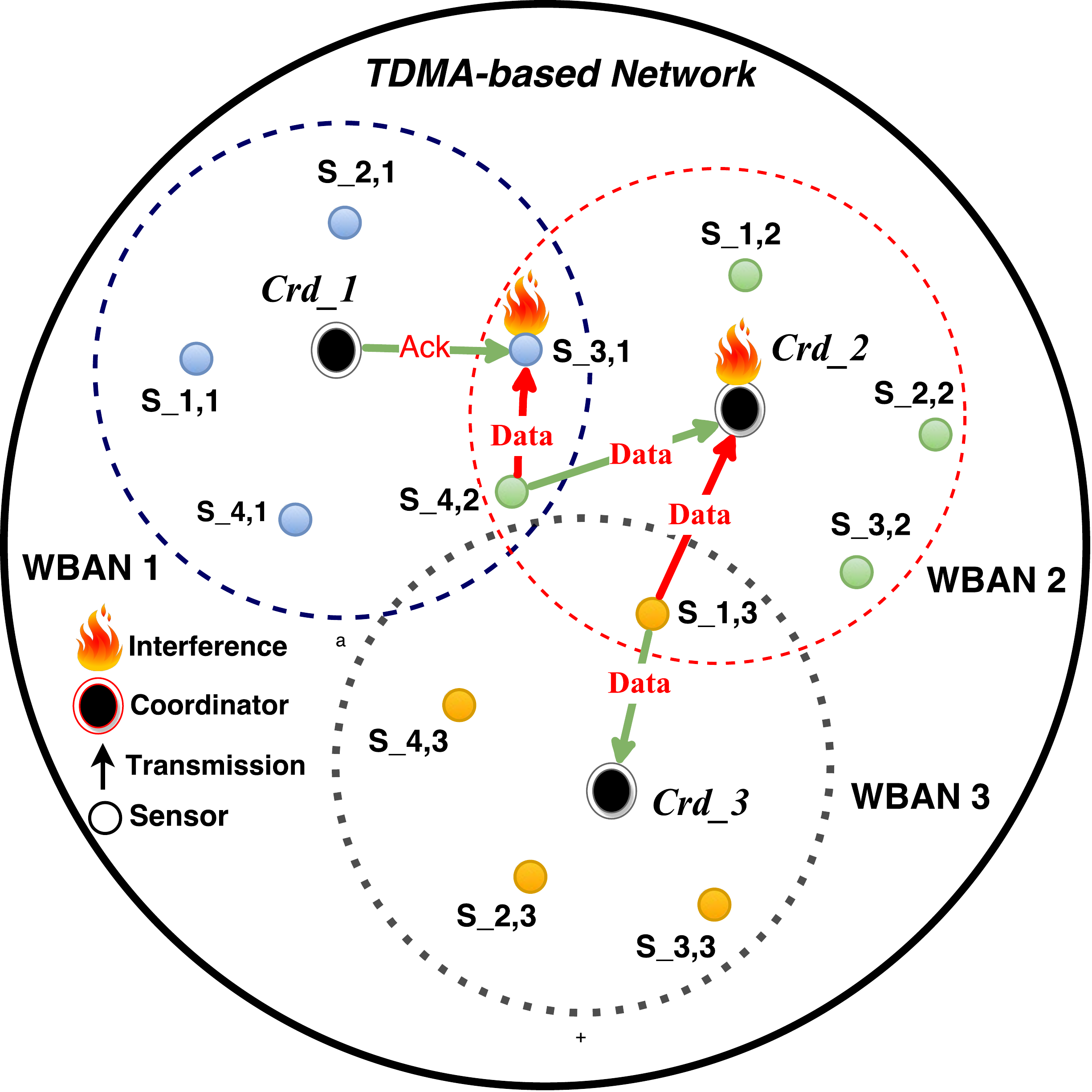}
\caption{Collision scenarios at sensor- and coordinator-levels}
\label{collision}
\end{figure}

In \textit{WBAN}s, data may be lost due to the co-channel interference, and hence acknowledgments are required to assure the transmitters the successful reception. Time-outs are used to detect reception failure at the corresponding receivers. We note that collisions may take place at the level of data or acknowledgement packets as shown in \textbf{Figure \ref{collision}} and explained below. 
\subsubsection{Data Packets Collision}
Data packet collisions take place at the coordinator when a sensor $S_{i,k}$ of $WBAN_k$ transmits while another sensor or coordinator of another $WBAN_q$ transmit on the same channel that $S_{i,k}$ uses, i.e., under the following condition: Coordinator of $WBAN_k$ is in range of $Crd_q$ or $S_{j,q}$ of $WBAN_q$, and $Crd_q$ or $S_{j,q}$ transmits on the same channel used by sensor $S_{i,k}$. In essence such collision may be experienced in two scenarios: (i)  $C_k = C_q$; i.e., both $WBAN_k$ and $WBAN_q$ happen to pick the same channel for intra-\textit{WBAN} communication, in which case $S_{j,q}$ or $Crd_q$ could be sending a data or acknowledgement packets, respectively. (ii)  Either $C_k$ = $BKC(S_{j,q})$ or $C_q = BKC(S_{i,k})$; i.e., the channel has been picked by $WBAN_k$ is equal to the same channel that has been allocated to sensor $S_{j,q}$ of $WBAN_q$ in its backup time-slot within the \textit{IMB} interference mitigation part.
\subsubsection{Acknowledgment Packets Collision}
Acknowledgment packet collisions take place at the sensor when a sensor $S_{i,k}$ of $WBAN_k$ receives while another sensor or coordinator of another $WBAN_q$ transmit on the same channel that $S_{i,k}$ uses, i.e., under the following condition: $S_{i,k}$ is in range of $Crd_q$ or $S_{j,q}$ of $WBAN_q$, and $Crd_q$ or $S_{j,q}$ transmits on the same channel used by sensor $S_{i,k}$. Similarly, the same condition of collision scenarios that are inferred by in data packets collision section still holds, i.e., (i) $C_k = C_q$ and, (ii) $C_k$ = $BKC(S_{j,q})$ or $C_q = BKC(S_{i,k})$. We argue that this case will be avoided by approach as explained later.
\subsection{Network Setup}
At the network setup time, each \textit{WBAN}’s coordinator will randomly pick a \textit{default operation channel} and a \textit{$M \times K$ Latin rectangle} from an orthogonal set. Initially, the coordinator instructs all sensors within its \textit{WBAN} to use the same \textit{default channel} along the whole \textit{TDMA} part. Meantime, the coordinator assigns a \textit{single symbol} from the symbol set \textit{\{{1,2,\dots,K}\}} to each sensor within its \textit{WBAN}, where the position hopping of each symbol in the Latin rectangle relates a \textit{single interference mitigation channel} and a \textit{unique backup time-slot}. Thereby, each coordinator determines the combination of a \textit{single interference mitigation channel} and a \textit{unique backup time-slot} for each sensor to eventually use in the \textit{IMB} part for interference mitigation. Subsequently, a coordinator informs each sensor within its \textit{WBAN} about its allocated: 1) \textit{interference mitigation channel} and, 2) \textit{backup time-slot} within the \textit{IMB} part of the superframe. Each coordinator reports this information to its sensors through beacon broadcast.
\subsection{Network Operation under \textit{CHIM}}
\textit{CHIM} depends on both acknowledgment and timeouts to detect collision/interference at both \textit{sensor-} and \textit{coordinator-} levels. In the \textit{TDMA} active part of a superframe, each sensor transmits a data packet in its assigned time-slot to the coordinator on the default operation channel, it sets a timeout timer and waits for an acknowledgment packet. Thus, if it successfully receives the acknowledgment packet from the corresponding coordinator, it considers the transmission successful, and hence it sleeps until the next superframe. In this case, the transmitting sensor does not need to switch to its allocated interference mitigation channel and use its dedicated backup time-slot in the \textit{IMB} part for interference mitigation. 

However, if the transmitting sensor does not successfully receive the acknowledgmenet within the time-out period, it assumes failed transmission due to interference and subsequently, it applies the interference mitigation procedure. Basically, the sensor waits until the \textit{TDMA} active part completes and then swtiches its channel to the allocated interference mitigation channel at the beginning of its allocated backup time-slot and retransmits its data packet. In fact, this failure is due to data or acknowledgment packets collisions at the \textit{coordinator-} or \textit{sensor-}levels, respectively, i.e., 1) the desired transmitted data packet is lost at the coordinator due to its interference from sensors in other \textit{WBAN}s at the same time or, 2) the acknowledgment packet of the desired coordinator is lost at the desired sensor due to the same reason. Therefore, depending on the acknowledgment packets and time-out period, both interfering sensors and coordinator address the collision problem in the same manner, each from its perspective.
\setlength{\textfloatsep}{1pt}
\begin{algorithm}
\footnotesize
\SetKwData{Left}{left}\SetKwData{This}{this}\SetKwData{Up}{up}
\SetKwFunction{Union}{Union}\SetKwFunction{FindCompress}{FindCompress}
\SetKwInOut{Input}{input}\SetKwInOut{Output}{output}

\Input{\textit{N} \textit{WBAN}s, \textit{K} Sensors/\textit{WBAN}, Orthogonal Latin rectangle \textit{OLR}}

\textit{Stage 1: Network Setup}

          \; \textbf{for} i = 1 \textbf{to} N
          
          \; \qquad $Crd_i$ randomly picks a single $DFC_i$ $\&$ $OLR_i$ for its $WBAN_i$
          
          \; \qquad \textbf{for} k = 1 \textbf{to} K
          
          \; \qquad \qquad $Crd_i$ allocates $BKC_{k,i}$ $\&$ $BKTS_{k,i}$ to $S_{k,i}$ from $OLR_i$

\textit{Stage 2: Sensor-level Interference Mitigation}

             \; \textbf{for} i = 1 \textbf{to} N
            
             \; \qquad \textbf{for} k = 1 \textbf{to} K
              
             \; \qquad \qquad $S_{k,i}$ transmits $Pkt_{k,i}$ in $TS_{k,i}$ to $Crd_{i}$ on $DFC_i$ in \textit{$TDMA_i$}
            
             \; \qquad \qquad \textbf{if} $Ack_{k,i}$ is successfully received by $S_{k,i}$ on $DFC_i$
            
             \; \qquad \qquad \quad $S_{k,i}$ switches to SLEEP mode until the next superframe
            
             \; \qquad \qquad \textbf{\textit{else}}
            
             \; \qquad \qquad \quad $S_{k,i}$ waits its designated $BKTS_{k,i}$ within $IMB_{i}$ part 
             
             \; \qquad \qquad \quad $S_{k,i}$ retransmits $Pkt_{k,i}$ in $BKTS_{k,i}$ to $Crd_{i}$ on $BKC_{k,i}$
            
\textit{Stage 3: Coordinator-level Interference Mitigation}     

             \; \textbf{for} i = 1 \textbf{to} N
            
             \; \qquad \textbf{for} k = 1 \textbf{to} K
            
             \; \qquad \qquad \textbf{\textit{if}} $Crd_{i}$ successfully received $Pkt_{k,i}$ in $TS_{k,i}$ on $DFC_i$
            
             \; \qquad \qquad \quad  $Crd_{i}$ transmits $Ack_{k,i}$ in $TS_{k,i}$ to $S_{k,i}$ on $DFC_i$
            
             \; \qquad \qquad \textbf{\textit{else}}
            
             \; \qquad \qquad \quad $Crd_{i}$ will tune to $BKC_{k,i}$ to receive from $S_{k,i}$ in $IMB_{i}$
             
             \; \qquad \qquad \quad $Crd_{i}$ receives $Pkt_{k,i}$ in $S_{k,i}$'s $BKTS_{k,i}$ on $BKC_{k,i}$
            
\caption{Proposed \textit{CHIM} Scheme}
\label{chim}
\end{algorithm}
\DecMargin{1em}
The orthogonality property of Latin rectangles avoids inter-\textit{WBAN} interference by allowing a $WBAN_i$, to have its unique channel allocation pattern that does not resemble the pattern of other \textit{WBAN}s.\/*, i.e., they do not share the same symbol positions, each in its own Latin rectangle and consequently, no other \textit{WBAN} in the network would simultaneously share the same pattern with $WBAN_i$ all the time. Generally, \textit{CHIM} makes it highly improbable for two transmissions to collide*/. Nonetheless, collision may still occur when (i) two \textit{WBAN}s randomly pick the same Latin rectangle, or (ii) more than \textit{16} \textit{WBAN}s coexist in the same area.\/*\textcolor{red}{, which means that, the number of \textit{WBAN}s exceeds the number of \textit{ZIGBEE} channels in the Latin rectangle.}*/ \textit{CHIM} handles these cases by optimizing the \/*\textcolor{red}{size and the reorganization of backup time-slots within the \textit{IMB} interference mitigation part, i.e., \textit{CHIM} may }*/reallocation of the backup channels and time-slots that are already allocated to interference-free sensors in the \textit{TDMA} data collection part to other interfering sensors. \textbf{Algorithm \ref{chim}} provides a summary of \textit{CHIM}.

\section{Mathematical Analysis}
In this section we opt to analytically assess the effectiveness of \textit{CHIM} in terms of reducing the probability of collisions. 
\subsection{\textit{TDMA} Collision Probability}
In this section, we derive the probability for a designated sensor that experiences collision within the \textit{TDMA} data collection part of the active frame. Let us consider a sensor $S_i$ of $WBAN_i$ that is surrounded by \textit{P} different sensors $S_j$, where $i \neq j$. For simplicity, we assume that $S_i$ transmits one data packet in a single time-slot within the \textit{TDMA} data collection part. $S_i$ successfully transmits its data packet on the default channel to the coordinator, \textit{iff}, none of the \textit{P} sensors transmits in the same time-slot using $WBAN_i$ default channel. Now, let \textit{X} denote the random variable representing the number of sensors that are transmitting their data packets in the same time-slot as $S_i$, if \textit{x} sensors transmit in the same time-slot of $S_i$, the probability of event \textit{X=x} is denoted by \textit{Pr(X=x)} and defined by \textbf{eq.} \ref{eq20} below.

\small
\begin{equation}\label{eq20}
         Pr\left(X=x\right)=C_{x}^{P}\alpha^x(1-\alpha)^{P-x}\left(min(M,K)/K\right)^{x},x\leq P
\end{equation}
\normalsize
Where $\alpha$ denotes the probability for a particular sensor $S_j$ of $WBAN_j$ to exist within the communication range of $WBAN_i$. Now, suppose \textit{Y} out of \textit{X} sensors schedule their transmissions according to Latin rectangles that are orthogonal to $WBAN_i$'s Latin rectangle, i.e., \textit{y} out of \textit{x} sensors select symbol patterns from other orthogonal Latin rectangles to $S_i$'s rectangle. Thus, the probability of \textit{y} sensors will not introduce any collision to $S_i$'s transmission is defined by \textbf{eq.} \ref{eq21} below. 

\small
\begin{equation}\label{eq21}
\begin{split}
 Pr\left(Y=y \mid X=x\right) &=\left(C_{y}^{K} C_{x-y}^{Z-K}\right)/C_{x}^{Z},\: x \leq P\: \&\: y\leq x
 \end{split}
\end{equation}
\normalsize
Where $Z = K\times m$ is the total number of symbol patterns in the orthogonal Latin rectangles family. However, \textit{X-Y} is a random variable representing the number of sensors that may collide with $S_i$'s transmission on the same channel; thus the probability that $S_i$'s transmission experiences collision is denoted by (\textit{collTx}) and defined by \textbf{eq.} \ref{eq22} below.

\small
\begin{equation}\label{eq22}
\begin{split}
Q &= Pr(collTx\mid Y=y, X=x)\\
  &=1-Pr(succTx \mid Y=y, X=x)\\
  &=1-\left((min(M,K)-1)/min(M,K)\right)^{x-y}\\
  &=1- \left(1 - 1/min(M,K)\right)^{x-y}\\
\end{split}
\end{equation}
\normalsize
Where \textit{Q} represents the probability that a sensor $S_i$ faces collision in one of its assigned time-slots and \textit{min(M,K)} represents all possible transmission time-slots for each $S_i$ within the \textit{TDMA} data collection part of the active frame. Thus, we depend on \textit{Q} to determine the whole number of sensors, denoted by \textit{W}, that face collisions within the \textit{TDMA} data collection part, where each sensor $S_i$ $\in$ \textit{W} will use its designated backup channel and time-slot within the \textit{IMB} interference mitigation part. Accordingly, we determine the new set of backup sensors that face collisions in the \textit{IMB} interference mitigation part in the following subsection.
\subsection{IMB Collision Probability}
In this subsection, we determine the probability of each backup sensor $S_i$ that faces collision in the \textit{IMB} interference mitigation part, when it uses its designated backup channel and time-slot. Let $T_{imb}$ denote the number of interfering sensors that collide both in the \textit{TDMA} data collection and the \textit{IMB} interference mitigation parts, where $T_{imb}$ follows binomial distribution. If \textit{t} sensors of a particular \textit{WBAN} face collision in the \textit{IMB} interference mitigation part, then the probability of event $T_{imb} = t$ is denoted by $Pr(T_{imb} = t)$ and defined by \textbf{eq.} \ref{eq55} below.

\small
\begin{equation}\label{eq55}
        Pr(T_{imb} = t) = C_{t}^{K} (Q^{2})^t (1-Q^{2})^{K-t}, \: t\leq K 
\end{equation}
\normalsize

And $Q^{2}$ is due to the 2-stage collision, i.e., the first collision happens in the \textit{TDMA} data collection part and the second happens in the \textit{IMB} interference mitigation part. Substituting Q of \textbf{eq.} \ref{eq22} in \textbf{eq.} \ref{eq55}.

\small
\begin{equation}\label{eq65}
        Pr(T_{imb} = t)=C_{t}^{K} (Q^{2})^{t} (1-Q^{2})^{K-t},\: t \leq K
\end{equation}
\normalsize
\small
\begin{equation}\label{eq100}
\begin{split}
 Pr(T_{imb}=t)&=C_{t}^{K}\times  (Q^{2})^{t} (1-Q^{2})^{K-t} ,\: t\leq K\\
&=C_{t}^{K} \times (1-1/min(M,K))^{(x-y)(K-t)}\\
&\times (2-(1-1/min(M,K))^{x-y})^{K-t}\\
&\times (1-(1-1/min(M,K)^{x-y}))^{2t}
\end{split}
\end{equation}
\normalsize
As a baseline for comparison, \textit{ZIGBEE} standard \cite{key26} shows that the active period of the superframe can be divided into two parts, \textit{TDMA} \textit{ZIGBEE} part and contention free period part (\textit{CFP}), where some sensors may require additional guaranteed time-slots (\textit{GTSs}) in the \textit{CFP} to avoid collisions have been experienced in the \textit{TDMA} \textit{ZIGBEE} part and complete their transmissions. However, these sensors use the same channel to transmit their pending data.  

\begin{table}

\centering
\caption{Simulation parameters}
\label{parm}
\begin{tabular}{ll}
%\noalign{\smallskip}
%\hline\noalign{\smallskip}
\hline\hline
 Sensor \textit{TxPower(dBm)} & -10\\\hline
 Sensors/\textit{WBAN} &20\\\hline
 \textit{WBAN}s/Network &Variable \\\hline
 Slots/\textit{TDMA} \textit{CHIM} part  &20\\\hline
 Slots/\textit{IMB CHIM} part  &20\\\hline
 Slots/\textit{TDMA} \textit{ZIGBEE} part  &20\\\hline
 Slots/\textit{CFP} \textit{ZIGBEE} part  &12\\\hline
 Latin Rectangle Size&$16 \times 20$\\\hline
%\noalign{\smallskip}\hline
%\hline\noalign{\smallskip}
\end{tabular}
\end{table}
\section{Performance Evaluation}
We have performed simulation experiments to validate the theoretical results and study the performance of the proposed \textit{CHIM} scheme. In this section, we compare the performance of \textit{CHIM} with \textit{ZIGBEE} standard \cite{key16}, which assigns guaranteed time-slots (\textit{GTSs}) in the \textit{CFP} to sensors that have experienced interference in \textit{ZIGBEE} \textit{TDMA} period of the superframe. The simulation parameters are provided in \textbf{Table \ref{parm}}.
\begin{figure*}
\begin{minipage}[b]{.3075\textwidth}
\centering
\includegraphics[width=1\textwidth, height=0.2\textheight]{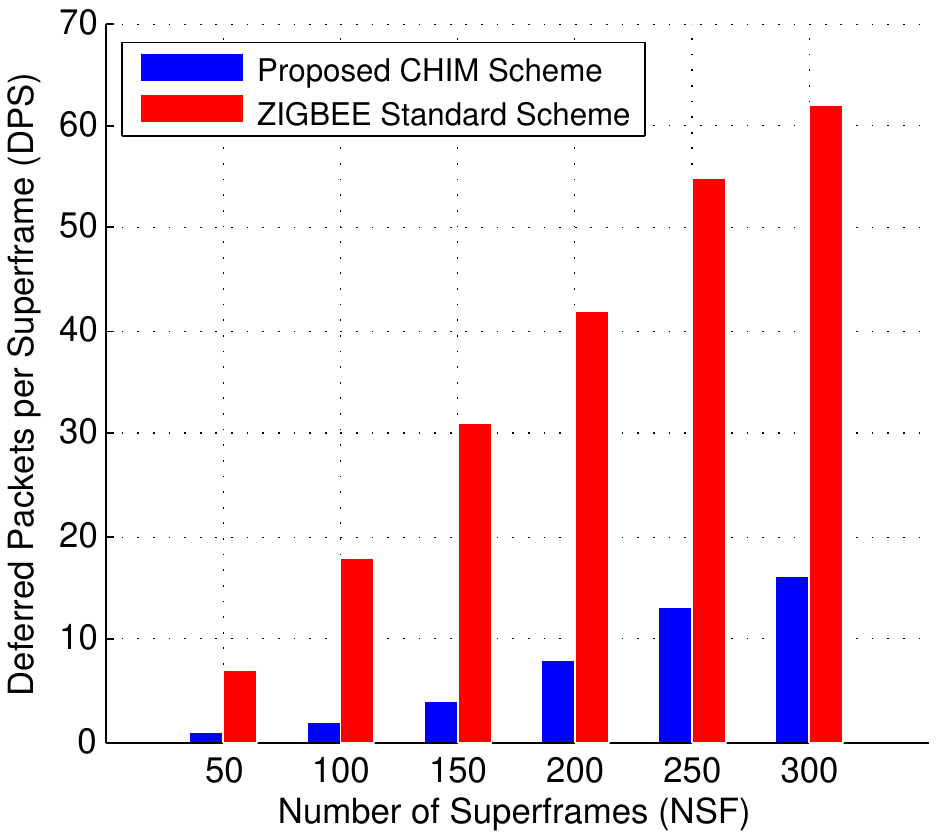}
\caption{Average number of deferred data frames (\textit{DPS}) versus the number of transmitted superframes (\textit{NSF})}
\label{delay}
\end{minipage}\qquad
\begin{minipage}[b]{.3075\textwidth}
\centering
\includegraphics[width=1\textwidth, height=0.2\textheight]{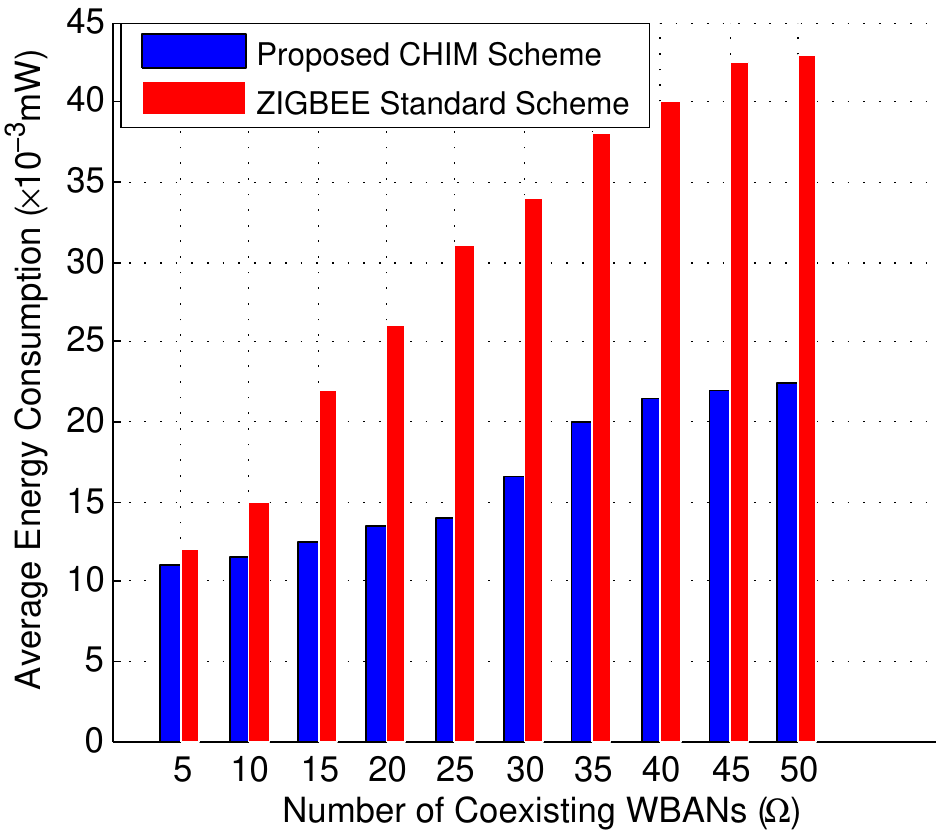}
\caption{Average energy consumption (\textit{AEC}) versus the number of coexisting \textit{WBAN}s ($\Omega$)}
\label{AEC}
\end{minipage}\qquad
\begin{minipage}[b]{.3075\textwidth}
\centering
\includegraphics[width=1\textwidth, height=0.2\textheight]{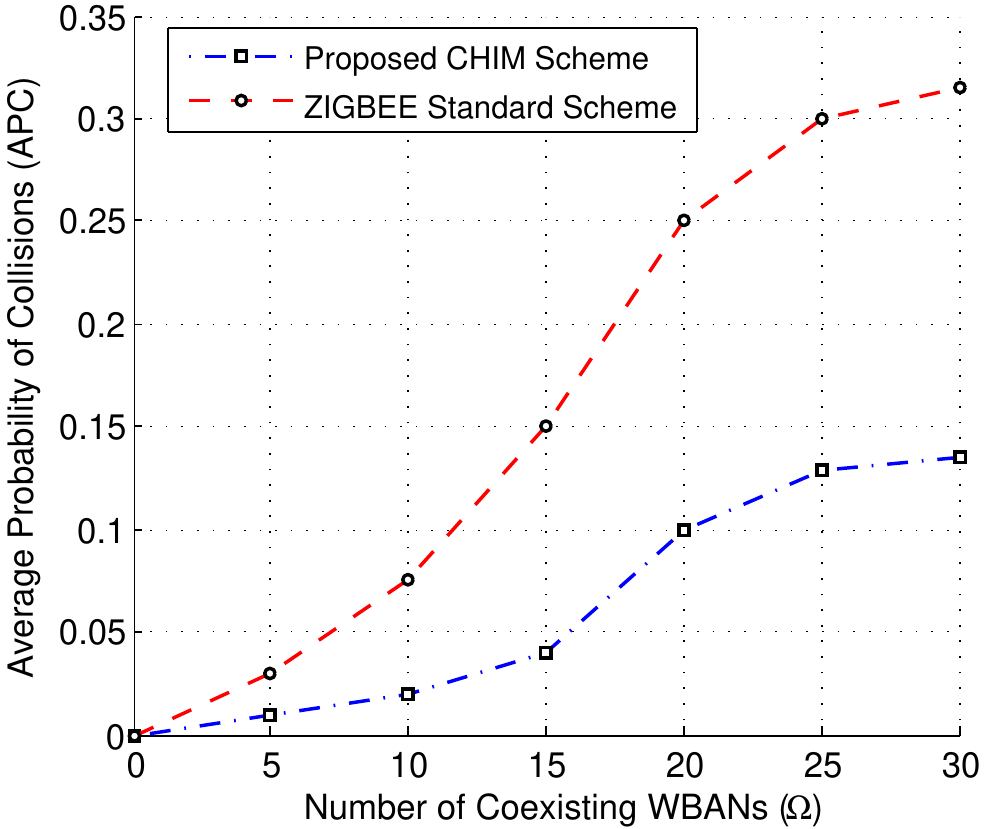}
\caption{\textit{WBAN} average probability of collisions (\textit{APC}) versus the number of coexisting \textit{WBAN}s ($\Omega$)}
\label{FOC}
\end{minipage}\qquad
\end{figure*}
\subsection{Frequency of Collisions}
The average probability of collisions denoted by \textit{APC} versus the number of coexisting \textit{WBAN}s ($\Omega$) for \textit{CHIM} and that for \textit{ZIGBEE} are compared in \textbf{Figure \ref{FOC}}. As can be clearly seen in the figure, \textit{CHIM} provides a much lower \textit{APC} because of the channel hoppings. It is observed from this figure that for \textit{CHIM} \textit{APC} is very low when $\Omega \leq 15$, which is due to the large number of channel hopping possibilities. When $15 < \Omega \leq 25$, \textit{APC} significantly increases due to the growth in the number of sensors which makes it possible for two or more sensors to be assigned the same channel in the same time slot. However, when $\Omega$ exceeds 25, \textit{APC} increases very slightly and eventually stabilizes at $135 \times 10^{-3}$ because of the maximal number of collisions is attained by each \textit{WBAN}. In \textit{ZIGBEE}, \textit{APC} slightly increases when $0 < \Omega \leq 10$, i.e., the number of interfering sensors and the number of available \textit{GTSs} are similar. Then, when $10 < \Omega \leq 25$, \textit{APC} significantly increases due to the growth in the number of interfering sensors which makes it possible for two or more sensors to collide in the same \textit{GTS}. \textit{APC} stabilizes at $315 \times 10^{-2}$ when $\Omega \geq 25$ reflecting maximum interference as all \textit{GTSs} are already assigned. %In addition, the simulation confirms the analytical results in terms of collision probability as shown in \textbf{Figure \ref{spoc}}
%\begin{figure}
%  \centering
%        \includegraphics[width=0.275\textwidth]{frequencyofcollision.pdf}
%\caption{\textit{WBAN} average  probability of collisions (\textit{APC}) versus the number of coexisting \textit{WBAN}s ($\Omega$)}
%\label{FOC}
%\end{figure}

\subsection{Energy Consumption}
\textbf{Figure \ref{AEC}} shows the average energy consumption of each \textit{WBAN} denoted by \textit{AEC} versus the number of coexisting \textit{WBAN}s ($\Omega$) for \textit{CHIM} and \textit{ZIGBEE}. As evident from \textbf{Figure \ref{AEC}}, \textit{AEC} for \textit{CHIM} is always lower than that of \textit{ZIGBEE} for all values of $\Omega$. Such distinct performance for \textit{CHIM} is mainly due to the reduced collisions that lead to fewer retransmissions and consequently lower power consumption. For \textit{CHIM}, the figure shows that \textit{AEC} slightly increases when $\Omega \leq 20$, i.e., there is a larger number of channel hopping possibilities than the interfering sensors which lowers the number of collisions among sensors and hence the energy consumption is decreased. \textit{AEC} significantly increases when $\Omega \leq 40$ due to the large number of sensors competing for the same channel in the same time-slots which results in more collisions and hence more energy consumption. When $\Omega$ exceeds 40, the energy consumption increases slightly to attain the maximum of $22.5 \times 10^{-3}mW$. However, in \textit{ZIGBEE}, \textit{AEC} slightly increases when $0 < \Omega \leq 10$, i.e., the number of interfering sensors and the number of available \textit{GTSs} are similar. Then, when $10 < \Omega \leq 40$, \textit{AEC} significantly increases due to the growth in the number of interfering sensors which makes it possible for two or more sensors to collide in the same \textit{GTS}. \textit{AEC} stabilizes at $43 \times 10^{-3}mW$ when $\Omega \geq 45$ reflecting maximum interference as all \textit{GTSs} are already assigned. 
%\begin{figure}
%\centering
%        \includegraphics[width=0.275\textwidth]{one.pdf}
%\caption{Average energy consumption (\textit{AEC}) versus the number of coexisting \textit{WBAN}s ($\Omega$)}
%\label{AEC}
%\end{figure}
\subsection{Transmission Delay}
The average number of deferred data packets of \textit{N = 20} coexisting \textit{WBANs} denoted by (\textit{DPS}) versus the number of transmitted superframes (\textit{NSF}) for \textit{CHIM} and \textit{ZIGBEE} are compared. \textbf{Figure \ref{delay}} shows that \textit{DPS} for \textit{CHIM} is always lower than that of \textit{ZIGBEE} for all values of \textit{NSF} which can be attributed to the reduced medium access contention that leads to fewer number of deferred data packets and consequently lower transmission delay. \textit{DPS} of \textit{ZIGBEE} is higher than that of \textit{CHIM} due to the usage of one instead of 16 channels and hence the number of competing sensors to the available \textit{GTSs} is large enough, which leads to larger number of deferred data packets and consequently longer transmission delay.  
%\begin{figure}
%\centering
%        \includegraphics[width=0.275\textwidth]{delay.pdf}
%\caption{Average number of deferred data frames (\textit{DPS}) versus the number of transmitted superframes (\textit{NSF})}
%\label{delay}
%\end{figure}
\section{Conclusions}
In this paper, we have presented \textit{CHIM}, a distributed \textit{TDMA}-based interference avoidance scheme for coexisting \textit{WBAN}s based on the properties of Latin rectangles. \textit{CHIM} enables predictable channel hoppings to minimize the probability of collisions among transmission of sensors in different coexisting \textit{WBAN}s. Accordingly, each coordinator autonomously picks an orthogonal Latin rectangle and assigns each individual sensor within its \textit{WBAN} a backup time-slot and channel to use if it experiences  interference. \textit{CHIM} depends on the special properties of Latin rectangles to minimize the probability of both time and channel matching among sensors in different \textit{WBAN}s, and consequently reduces the transmission delay and energy consumption. Compared with competing schemes, \textit{CHIM} has low complexity and does not require any inter-WBAN coordination. We have analyzed the expected collision probability in the network and validated the advantages of \textit{CHIM} through simulation. The simulation results have shown that \textit{CHIM} achieves 55\% improvement in collision probability, 40\% in energy consumption and 30\% transmission delay.
%\bibliography{biblio}{}
%\bibliographystyle{plain}

\end{document}